# Planar narrow-band-pass filter based on Si resonant metasurface

Ze Zheng,[1,2,3] Andrei Komar,[1] Khosro Zangeneh Kamali,[1] John Noble,[4] Lachlan Whichello,[4] Andrey E. Miroshnichenko,[5] Mohsen Rahmani,[1,2] Dragomir N. Neshev,[1] and Lei Xu[1,2,a]

[1] *Research School of Physics, ARC Centre of Excellence for Transformative Meta-Optical Systems (TMOS), The Australian National University, Canberra, ACT 2601, Australia*
[2] *Advanced Optics & Photonics Laboratory, Department of Engineering, School of Science & Technology, Nottingham Trent University, Nottingham NG11 8NS, UK*
[3] *School of Physics, Nankai University, Tianjin 300071, China*
[4] *Seeing Macines Inc., Canberra ACT 2600, Australia*
[5] *School of Engineering and Information Technology, University of New South Wales, Canberra ACT 2600, Australia*

(Dated: 19 December 2020)

Optically resonant dielectric metasurfaces offer unique capability to fully control the wavefront, polarisation, intensity or spectral content of light based on the excitation and interference of different electric and magnetic Mie multipolar resonances. Recent advances of the wide accessibility in the nanofabrication and nanotechnologies have led to a surge in the research field of high-quality functional optical metasurfaces which can potentially replace or even outperform conventional optical components with ultra-thin feature. Replacing conventional optical filtering components with metasurface technology offers remarkable advantages including lower integration cost, ultra-thin compact configuration, easy combination with multiple functions and less restriction on materials. Here we propose and experimentally demonstrate a planar narrow-band-pass filter based on the optical dielectric metasurface composed of Si nanoresonators in array. A broadband transmission spectral valley (around 200 nm) has been realised by combining electric and magnetic dipole resonances adjacent to each other. Meanwhile, we obtain a narrow-band transmission peak by exciting a high-quality leaky mode which is formed by partially breaking a bound state in the continuum generated by the collective longitudinal magnetic dipole resonances in the metasurface. Our proposed metasurface-based filter shows a stable performance for oblique light incidence with small angles (within 10 deg). Our work imply many potential applications of nanoscale photonics devices such as displays, spectroscopy, etc.

## I. INTRODUCTION

Optical metasurfaces are planarized, ultra-thin, patterned artificial interfaces composed by subwavelength resonators, so-called optical meta-atoms, with structured electric and magnetic fields that can be used to shape light in exotic ways based on the excitation of plasmonic resonances[1–3] or high refractive index low-loss Mie-type resonances[4–7]. Although surface plasmons supported by plasmonic nanostructures are able to provide strong light-matter interactions at the nanoscale, these nanostructures suffer from the inherently high Ohmic losses of metals in the optical spectral region and they are easy to melt or deform when illuminated by highly intense lasers, therefore, the performances of conventional metal-based nanoparticles are limited. Nanostructures with high refractive index dielectric materials, such as Si, GaAs, GaP, etc, are transparent building blocks and have been demonstrated that they achieve unconventional electromagnetic responses of light. Such dielectric nanoparticles support both lower and higher-order optically-induced electric and magnetic Mie resonances[4–7]. Such a multipolar resonance feature of optical meta-atoms provides a powerful platform to achieve desired interference of different modes within a single meta-atom, yielding unique possibilities for an efficient control over the light scattering properties as well as the near-field distributions at the nanoscale. The generation and interference of multipolar resonances from dielectric nanostructures play more and more important role in modern nanophotonics, and have caused a surge in the research field of dielectric metaoptics.

Various interesting optical effects and phenomena have been discovered involving the interference between multipolar resonances in recent years, e.g., Kerker scattering where the interference between electric and magnetic dipole resonances leads to the suppression of light backscattering and realization of total transmission feature of light[8,9]; electric anapole state where the interference between the Cartesian electric dipole and the toroidal dipole modes cancel their far-field radiations and achieve a non-radiating state with enhanced near-fields[10]; magnetic anapole state where the interference between the Cartesian magnetic dipole and the mean square radii - a third-order term in the expansion of the magnetic dipole moment, which shares the same radiation pattern as the Cartesian magnetic dipole, further cancels their scatterings in the far-field domain[11,12]; Fano-type interference where a broadband "bright" resonance which is accessible from a free space, spatially and spectrally overlaps with a narrow-band "dark" resonance which is less-accessible or even inaccessible from the free space, and results in a narrow reflection or transmission window[13–15]. Recently, the concept of bound state in the continuum (BIC) supported by dielectric nanostructure has drawn extensive attentions due to its robust ability to confine light and tailor the radiative quality factor[16–19].

In recent years, employing the multipolar resonance feature, optical dielectric metasurfaces have led to a variety of applications in flat optical elements such as flat lenses[20–22], beam deflectors[23–26], holograms[27], quantum sources and quantum entanglement control devices[28–30] to name a few. As compared to the conventional bulky refraction-based optical elements, optical metasurfaces are of subwavelength thickness, which enable their easy integration into compact devices and realisation of miniaturized on-chip photonics elements. Importantly, as artificial 2-dimensional materials made of subwavelength inclusions, optical metasurfaces are not restricted to material compositions in order to achieve the desired optical functions, thus they can provide unparalleled steering over the different light properties.

---

[a] Electronic mail: lei.xu@ntu.ac.uk

Optical filter is a critical element for optical information technologies and light characterization. Optical filters can selectively allow light with certain wavelength ranges to be transmitted. It has been widely used in various applications, including spectral imaging[31,32], optical data storage[33], generation of security images[34], energy-efficient displays[35–37]. The most common approach to design an optical filter is based on the Fabry-Pérot (FP) resonance, in which the properties are defined by adjusting the refractive index and the light path[38–40]. Conventionally, filters are designed based on a FP resonator using a pair of broadband high reflective mirrors. By controlling the FP cavity thickness, optical filters with different center wavelengths can be generated. However, based on the working principle, FP filters are generally sensitive to the incident angle of light. Different incident angles lead to the different effective light paths in the FP cavity and thus result in a considerable shift of the center wavelength. Furthermore, the thickness of the filter is generally in the micrometer scale, which results in a greater optical distance for the occurrence of optical cross-talk. This will potentially hinder the integration of such optical filters into the on-chip nanophotonics devices. The development of nanofabrication and nanomanufacturing gives new opportunities for designing optical filters based on nanoplasmonics or dielectric Mie resonances[41–48]. In this paper, we employ the multipolar resonance feature of dielectric metasurfaces, combining it with the BIC mechanisms, and design a Si metasurface based planar narrow-band-pass (PNBP) filter in the near infrared (NIR) spectral region. The central wavelength can be controlled by the geometry of the Si resonators. In addition, owing to the mechanism of BIC, the bandwidth can be flexibly tuned via the degree of the geometric asymmetry of the Si resonators. We examined the angular dependence of the filter both theoretically and experimentally. As compared to the FB-cavity based filters, the proposed filter exhibits a better angular performance.

## II. RESULTS AND DISCUSSIONS

**Design of metasurfaces.** To realise a metasurface-based PNBP filter, there are two main points to be addressed: i) the realisation of a broadband transmission valley spectral region which blocks light for a given wavelength range, and ii) the realisation of a narrow-band transmission peak region which permit light transmission at specific desired wavelength. In the following, we first describe the working principle of our approach to realise the PNBP filter working in the NIR re-gion based on the multipolar resonance feature of dielectric metasurfaces. We consider an optical metasurface that consist of amorphous Si nanodisks on a square lattice with period $\Lambda = 650$ nm, height $h = 450$ nm on a glass substrate.

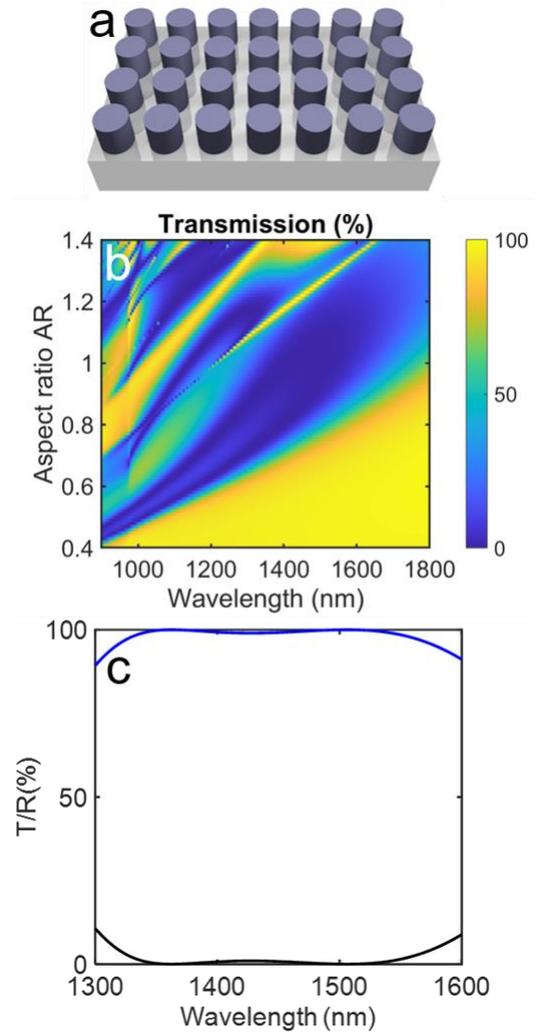

FIG. 1. (a) Schematics of the Si nanodisk metasurface. (b) Calculated transmission spectra as a function of the aspect ratio AR and light wavelength $\lambda$ . (c) Calculated transmission and reflection spectra for the Si nanodisk metasurface when $AR = 1.05$.

It is well known that exciting a resonance from a metasurface usually leads to a transmission dip in the spectrum based on the scattering from nanoparticles principle[49]. In order to obtain a broadband region with suppressed light transmission, one simple and direct approach is to combine multiple electric/magnetic resonances adjacent to each other spectrally in the desired wavelength range. In this case, when light illuminates on the metasurface, the transmission will be suppressed in a broadband spectral region based on the excitation of these resonances. For Mie resonances supported by the high-refractive-index dielectric nanoresonators, the width of the resonance usually decreases with the increase of the multipole's polar order $l$[49]. The lowest order dipoles, elec-tric dipole (ED) and magnetic dipole (MD), have the largest spectral widths (around 100 nm in the NIR range). In the following, we employ ED and MD resonances supported by Si nanodisks (Figure 1). By tuning the spectral position of ED and MD adjacent to each other resonances, we have realised a broadband transmission valley of around 200 nm. The relative spectral position of ED and MD can be adjusted by tuning the aspect ratio of the nanodisks, $AR = D/h$, where $D$ and $h$ are the diameter and the height of the nanodisks. Figure 1(b) shows the calculated transmittance as a function of the aspect ratio $AR$ and wavelength. As can be seen, via varying the aspect ratio $AR$, the ED and MD resonances can be tuned spectrally and merge together. Specifically, when $AR = 1.05$, the ED and MD resonances form a wide spectral region around 200 nm with nearly total suppression of light transmission through the metasurface, as shown in Figure 1(c).

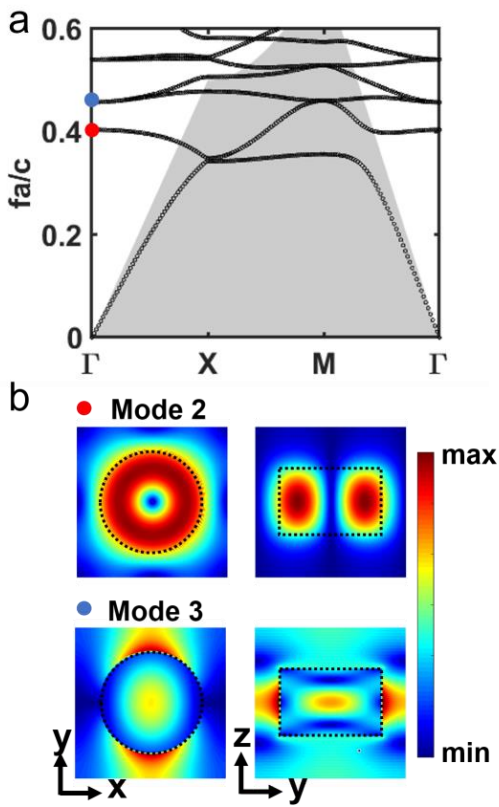

FIG. 2. (a) Calculated band structure for a periodic 2D array of Si nanodisks of radius 230 nm. The gray shading indicates the area located below the free-space light cone. The red and blue dots mark the two modes at the Γ point. (b) Electric near-field distributions of the Mod 2 (red dot) and Mode 3 (blue dot), respectively.

To realise a narrow-band peak out of the transmission valley produced by the ED and MD resonances, we employ a longitudinal MD resonance supported by the metasurface. This resonance cannot be excited by normal light illumination due to symmetry mismatch[18]. The MD resonance belongs to the symmetry-protected BICs in sub-diffractive periodic system. By introducing geometric asymmetry, such as an air hole in the nanodisks, it allows opening a radiation channel and transform this ideal-BIC MD state into a quasi-BIC MD state with the finite Q-factor corresponding to a tuneable bandwidth based on the degree of the geometric asymmetry. At Γ point of the band structure of the optical metasurface when the operating frequencies are below the diffraction limit, the only radiating states are plane waves in the normal direction with the electromagnetic field being odd under $180^o$ rotation around z-axis, i.e., $C_2$ symmetry, thus any even mode at the Γ point is a BIC due to the zero overlap between their mode profiles and outgoing waves[17]. As an example shown in Figure 2(b), Mode 2 is an even mode, while Mode 3 is an odd mode. Both of them sit above the free space light cone, while only Mode 2 is a symmetry-protected BIC. Such a BIC can be realised as a quasi-BIC by introducing the asymmetric factors into the system, and it will manifest a strong resonance with the high Q-factor in the transmission spectrum[18,41,50–52]. Here we employ this quasi-BIC to produce the narrow-band-pass feature for the filtering purposes. The relative spectral position of our resonances (ED, MD and quasi-BIC) can be adjusted accordingly via the geometric parameters of the metasurface, transverse size and height of the resonators. It is worth noting that elliptical-shaped resonators can also be employed to adjust the relative spectral position of the central wavelength for the narrow-band peak with respect to the transmission valley region.

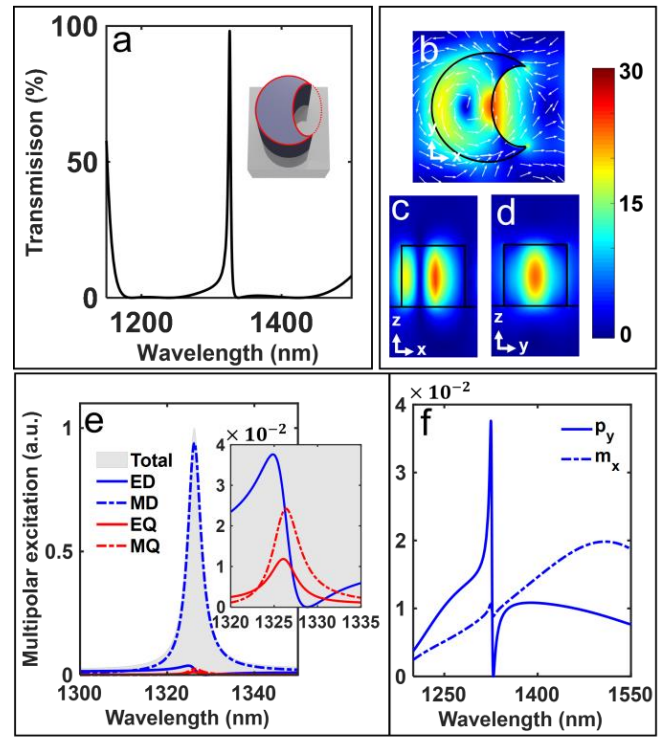

FIG. 3. (a) Calculated transmission spectrum of the Si disk-hole metasurface. Inset: schematic of the unit cell of our Si disk-hole metasurface. (b) Electric near-field distributions at the wavelength of 1329 nm. (c) Spherical multipolar structure of the Si disk-hole metasurface. (d) Cartesian ED $p_y$ and MD $m_x$ modes excitations.

**Multipolar structure.** Figure 3(a) shows the calculated transmission spectrum of our designed Si resonant metasurface. As can be seen, a narrow-band peak around 1329 nm emerges out of the broadband transmission valley in the spectrum, indicating the realisation of a filtering functionality based on the Si resonant metasurface. Due to the decrease of the total volume size of the Si nanoresonators after the air hole is introduced, one sees that the resonant positions for the transmission valley and peak have shifted together towards the shorter wavelength direction as compared to Figure 1. For an inverse design of the large number of filters with various central wavelengths, it is worth noting that the machine learning approach can be applied, which can introduce the remarkable design flexibility, that can exceed performance of the conventional optimisation methods for such types of the inverse design problems in nanophotonics[53–56]. Circular displacement currents are excited inside the Si nanodisk, as depicted by the electric near-field images shown in Figure 3(b-d), indicating the excitation of the longitudinal MD resonance. We also performed the multipolar analysis, Figure 3(e,f) show the multipolar structure of the metasurface. As can be seen, the optical response is dominated by the designed MD excitation around 1329 nm, with a small contribution from the ED, EQ and MQ excitations. Due to the $C_2$ symmetry of our subdiffractive system, the MD, EQ and MQ do not contribute to the far-field radiation of our metasurface. In other words, here MD, EQ and MQ form the dark channel which cannot be excited by the external incoming light, and the ED forms the leaky channel which couples the incoming light into the dark channel and also out-couples the light into the far-field domain. This interference between the two levels leads to the narrow transparency peak in the transmission spectrum. Figure 3(f) shows the excitation of the electric dipole $p_y$ and the magnetic dipole $m_x$. This corresponds to the form of the transmission valley. A sharp Fano-shape feature can be clearly observed from the excitation of $p_y$, i.e., the leaky channel, which is formed near the introduced air hole in the nanodisk. The presence of this $p_y$ further causes the assymetic electric near-field distribution as shown in Figure 3(b-d).

**Angular dependence.** We further investigate the performance of our metasurface based PNBP filter with different input beam angles, the angle of the propagation direction to $yz$ plane $\theta$, and the angle of the propagation direction to $xz$ plane $\theta_2$ (Figure 4(a)). As shown in Figure 4(b) and 4(c), our simulation result shows that for $\theta$ and $\theta_2$ less than 10 deg the proposed PNBP filter can maintain a stable performance. For comparison, we also calculated the angular dependence of the transmission spectra from a FP filter as shown in Figure 4(d). Here, the refractive index is set to be $n_1=1$ and $n_2=1.56$. The distance between the two surfaces of the etalon is set to be $d=2.55$ μm in order to generate a FP resonance with central wavelength around 1330 nm. As can be seen, for the FP filter, when the incident angle increases from 0 deg to 10 deg, the central wavelength has shifted around 20 nm. For the metasurface based PNBP filter, the shift of the central wavelength is around 3 nm, which shows a better and more stable performance when the input light angle varies.

**Experimental characterisation.** To verify our theoretical and computational results experimentally, we fabricated the Si nanodisk metasurface on a glass substrate. Figure 5(a) shows the scanning electron microscope (SEM) image of the fabricated sample. The sample consists of the Si nanodisks with hole defects arranged in a periodic order. A concave surface curvature in the vertical direction is observed from the fabricated sample that decreases the total volume of the Si nanodisks. The effect of this fabrication imperfections will be discussed in the following. We investigated the performance of our sample by measuring the transmission spectrum under different angles of incident light. We used a home-built microscope setup with a NIR spectrometer (Ocean Optics NIRQUEST) connected. The setup consists of a source of light, a halogen lamp, a polarizer, to control the polarization of the incident light, a Köhler illumination system, to homogeneously illuminate the sample and control the $k$-vector of incident light, the sample on the holder, which allows to rotate it, to measure the transmission spectra for different angles of incidence, a microscope objective (Mitutoyo×20 NA = 0.4), and an additional aperture placed in a confocal system to cut the background radiation at the sample image plane. The transmission spectra were referenced to the transmittance of the glass substrate next to the metasurface. As shown in Figure 5(b), under y-polarised light illumination, the metasurface demonstrates a narrow transmission peak out of a broadband transmission valley region. A spectral shift of the sample is observed in our experiment as compared to our previous theoretical results, possibly due to the above-mentioned fabrication imperfections which causes a reduce in the volume size of the Si nanodisks. Despite of the spectral shift caused by the fabrication imperfections, our experimental results successfully demonstrate our theoretical approach. The robustness of this type of BIC to the shape irregularities of the individual nanodisks makes it a promising candidate for various filtering applications[57]. Figure 5(c,d) show the experimentally measured angular dependence of the transmission spectra of our sample. As can be seen, the angular transmittance of our fabricated filter is lower than for a FB cavity filters and is in a good agreement with our previously simulated results shown in Figure 4.

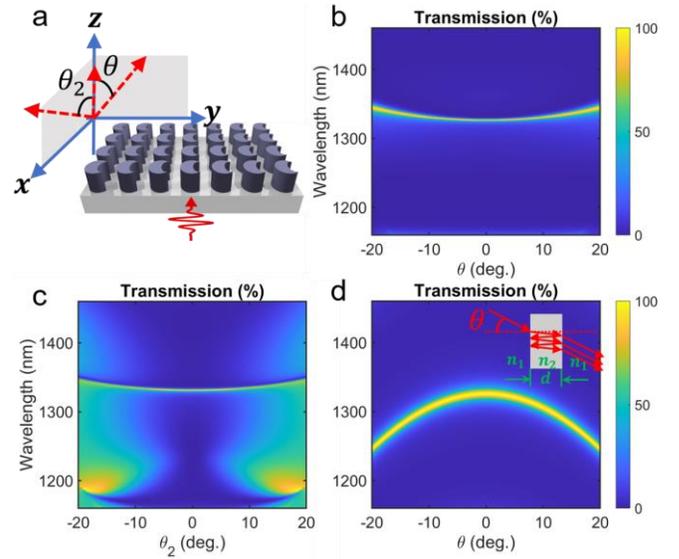

FIG. 4. (a) Top: Schematics of the Si disk-hole metasurface under light incidence with different angle $\theta$ or $\theta_2$. (b) Calculated transmission spectra under different light incidence angle $\theta$. (c) Calculated transmission spectra under different light incidence angle $\theta_2$. (d) The transmission spectra for a conventional FB filter under different light incidence angle $\theta$. The inset shows the schematic of a simplified FP cavity.

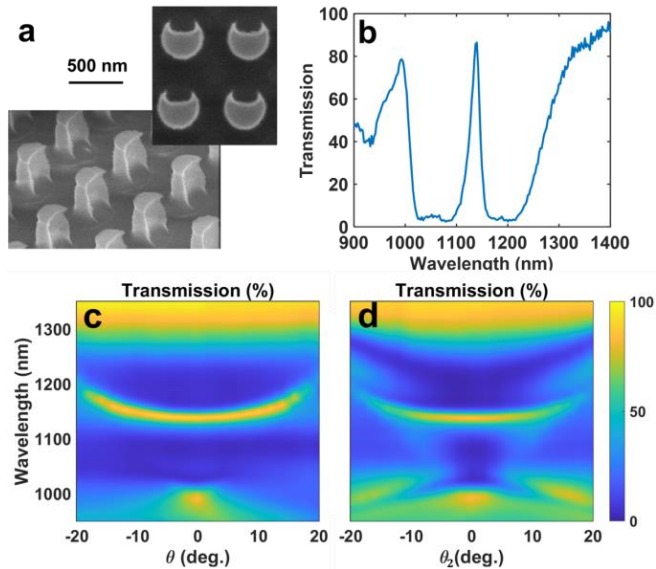

FIG. 5. (a) The SEM image of the fabricated Si disk-hole metasurface sample. (b) Experimentally measured transmission spectrum under normal light incidence. (c) Experimentally measured transmission spectra under different light incidence angle $\theta$. (d) Experimentally measured transmission spectra under different light incidence angle $\theta_2$.

## III. CONCLUSION

In summary, we have proposed a planar narrow-band-pass filter based on Si resonant metasurface, which demonstrates almost flat angular response in the range of $\pm 10°$ angles of incidence. Our approach is based on the multipolar resonance feature of optically resonant nanostructures. By properly tuning the spectral positions of ED and MD resonances, we have obtained a broadband transmission valley from the metasurface. A longitudinal MD resonance supported by the Si nanoresonators is used to realise the narrow transmission peak region. This longitudinal MD resonance is originated from the symmetry protected BIC mechanism supported in the subdiffractive periodic systems. The width of the transmission line can be easily controlled by the degree of the geometric symmetry mismatch. Our results may have implications for nanophotonics applications including displays, spectral imaging, etc. Furthermore, our approach combines the multipolar feature of meta-optics and the BIC mechanism, to control and engineer the coupling between nanostructures and free-space electromagnetic waves. The approach introduces special functionalities in optical metasurfaces, and thus, might be instructive for the design of high-performance metadevices.

## ACKNOWLEDGMENTS

The authors acknowledge the funding support provided by the ARC Discovery Project (DP200101353) and ARC Linkage Project (LP180100904). M.R. acknowledges support from the Royal Society and the Wolfson Foundation. A.E.M. appreciates the support by UNSW Scientia Fellowship. The authors appreciate the use of the Australian National Fabrication Facility (ANFF) – the ACT Node.